\documentclass[,final ]{aipproc}
\layoutstyle{6x9}
 \usepackage{graphicx}
 \usepackage{psfrag}
\usepackage{epsfig}
\def\be{\begin{eqnarray} &&}

\def\ee{\end{eqnarray}}
\def\psla{\rlap \slash}

\begin{document}
\title{The Electromagnetic Form Factor of the Kaon in the Light-Front Approach}
%% \texttt{http://www.aip..org/pacs/index.html}>}
\author{Edson~O.~da~Silva, J.~P.~B.~C.~de~Melo, Victo~S.~Filho~and~Bruno~El-Bennich}
{
address={
Laborat\'orio de F\'\i sica Te\'orica e Computacional, LFTC, 
Universidade Cruzeiro do Sul, 01506-000,~S\~ao Paulo, Brazil 
}}
\begin{abstract}
The kaon electromagnetic form factor 
is calculated within a light-front constituent quark model 
(LFCQM). 
The electromagnetic components of the current are extracted from 
the Feynman triangle diagram within the light-front approach. 
We also obtain the electroweak decay constant and the charge radius for the kaon in the light-front approach.
In this work, the kaon observables are calculated and a fairly good agreement is obtained with 
a very higher accuracy when compared with the experimental data.
\end{abstract}
\maketitle

%% \section{Introduction}
With the light-front constituent quark model,~{\it LFCQM}, it 
is possible to describe hadronic physics, in terms of {\it QCD} freedom degrees~\cite{Brodsky98}. 
In this work, we describe 
in a consistent way the hadronic bound state composite system 
$q\overline{q}$ (the kaon meson) and the corresponding 
electromagnetic (e.m.) form factor in the light-front model. In the last years, the pion e.m. form 
factor has been calculated in many works (for instance, see Refs. ~\cite{Pacheco2006,Tsirova,Leitner11} and 
references therein); but for the kaon, there are few theoretical results~\cite{Maris1997,Pereira,Raha2010}. 
The extraction of the electromagnetic form factor in the 
light-front approach depends on which component of the 
electromagnetic current~(e.c.) is utilized. This is related to problems with the rotational 
symmetry breaking~\cite{Pacheco98,Pacheco99,Naus98,Pacheco97,Ji2001}.  
The e.c. with the light-front approach has another contribution, besides 
the valence contribution to the electromagnetic current. 
That contribution corresponds to the pair terms added to the 
matrix elements of the e.c.~\cite{Pacheco98,Naus98,Pacheco2002}. 
In this work, we report results for the kaon e.m. form factor 
which is extracted from the positive component of the 
e.c. in the light-front formalism, $J_K^+= J^0 + J^3$, with a pseudoscalar 
coupling of the quarks in the Breit frame 
($q^+=0$, $q_{\perp}=(q_x,0) \ne 0$). 
In the case of $J_K^+$, there is no pair-term contribution in this frame. 
However, for the $J_K^-$ component of 
the e.c., the pair term contribution is 
diferent from zero and necessary in order to preserve the 
rotational symmetry of the current.
Then, the matrix elements of the e.c. 
within the light-front approach receive other contribution, 
besides the valence one~(see Refs.~\cite{Pereira,Pacheco99,Ji2001} and 
 references therein). 

Studies concerning light vector and scalar mesons are important 
because they help our understanding of 
QCD in the non-perturbative regime. 
 
The kaon e.m. form factor is calculated from the 
Feynman triangle diagram in the impulse aproximation. 
The e.c.~$J^{\mu}$ can be writen
in terms of the charge and quark fields~$q_f$ as:
\begin{eqnarray}
J_q^\mu(q^2) &=&-\imath 2 e_q \frac{m^2}{f^2_K}
N_c\int \frac{d^4k}{(2\pi)^4} {\mathrm{Tr}} \Bigl[ S(k)\gamma^5 S(k-p^{\prime})
\gamma^\mu S(k-p) \gamma^5 \Bigr]\Gamma(k,p^{\prime})\Gamma(k,p) 
\label{jcurrent}         
\ \ , 
\nonumber \\ 
\nonumber    \\
J^\mu_{\bar{q}}(q^2) & = & q \leftrightarrow \bar {q} \ \   
\mbox{in} \ \ J^{\mu}(q^2)  \ \ , 
\end{eqnarray} 
in which $N_c=3$ is the numbers of colors, 
 $e_q$ is the quark (anti-quark) charge and 
 $S(p,m)=1/(\psla{p}-m^2 + \imath \epsilon)$ are 
 the fermion propagators. 
The calculations are performed in 
the Breit frame, ($q^{+}=0$), 
with $p^{\mu}=(0,-q/2,0,0)$ and $p^{\prime {\mu}}=(0,q/2,0,0)$, 
for the initial and 
final momenta of the system, respectively. The momentum transfer is 
$q^{\mu}=(0,q,0,0)$ and $k^{\mu}$
is the spectator quark momentum. The 
factor 2 appears due to the isospin algebra.  
The function $\Gamma(k,p)$ is the regulator vertex function, 
used in order to regularize the Feynman triangle 
diagram of Eq.(\ref{jcurrent}). 
We have used as $q\bar{q}$ vertex function 
the nonsymmetric vertex~~~$\Gamma^{NSY}(k,p)=
N/((p-k)^2-m^2_R+\imath\epsilon)$~\cite{Pacheco99}.
The matrix elements of the e.c.~for the kaon
yields the e.m.~form factor 
\begin{equation}
(p+p^{\prime})^{\mu} 
F_{K}(Q^2)\ = \ <K(p^{\prime})|J^{\mu}|K(p)>, 
\label{covariant}
\end{equation}
where $K$ is the kaon field operator and $Q^2=-q^2$.
The covariant kaon form factor 
is obtained from Eqs.~(\ref{jcurrent}) and 
(\ref{covariant}): 

\begin{equation}
F_{K}(q^2) = -\frac{i2eN_c}{(2P)} \frac{m^2}{f_{K^2}}
\int \frac{d^4 \,k}{(2\pi)^4}\,
Tr\left[S(k)\gamma^{5}S(P'-k)\gamma^{\mu}S(P-k)
\gamma^5 \right]    
\Gamma(P',k)\Gamma(P,k)  \ . 
\label{ffactor}
\end{equation}
The $J_K^{+}$ component of the e.c.~is
employed to 
extract the e.m.~form factor. 
%% The normalization constant $N$ it is found with  $F_{K}(0)=1$ for the kaon e.m.~form factor. 
The trace in Eq.(\ref{ffactor}) is the sum of the 
two contributions
\begin{eqnarray}
{\mathrm{Tr}}[1] &=& \left[\gamma^{5}
(k\!\!\!\slash - P\!\!\!\!\slash + m_u)\gamma^{\mu}
(k\!\!\!\slash - P'\!\!\!\!\!\slash + m_u)
\gamma^5 (k\!\!\!\slash + m_{\overline{s}}) 
\right] \ \ ,  \nonumber  \\ 
{\mathrm{Tr}}[2] &=& \left[\gamma^{5}
(k\!\!\!\slash - P\!\!\!\!\slash + m_{\overline{s}})\gamma^{\mu}
(k\!\!\!\slash - P'\!\!\!\!\!\slash + m_{\overline{s}})
\gamma^5 (k\!\!\!\slash + m_u) 
\right] \ \ .
\end{eqnarray}
The complete trace is 
%% \begin{equation}
${\mathrm{Tr}}_{\rm kaon}[...] =(2/3)~{\mathrm{Tr}}[1] +(1/3)~{\mathrm{Tr}}[2]$,
%%\label{resultant}
%% \end{equation} 
where the factors $2/3$ and $1/3$ are isospin factors. 
The integration in 
Eq.(\ref{ffactor}) has contributions from two intervals: 
$0<k^+<P^+$ and $P^+<k^+<P^{\prime +}$, 
where $P^{\prime +}=P^+ + \delta^+$. The first interval is the
contribution of the valence wave function to 
the e.m.~form factor and the second interval 
corresponds to the pair-term contribution~\cite{Pacheco98,Pacheco99,Naus98,Pacheco2002,Pacheco992}. 
In the case of the 
nonsymmetric vertex with the plus component
of the e.c., the second interval does not contribute at all~\cite{Pereira,Pacheco99}. 
One can verify that only the on-shell pole 
$\bar{k}^-=\frac{f_1-\imath \epsilon}{k^+}$ contributes 
to the $k^-$ integration in the interval $0<k^+<P^+$.  
 Hence, after Cauchy integration in the light-front 
energy $k^-$, Eq.~(3) becomes:
\begin{eqnarray}
F^{+}_{\bar{q}}(q^2)
&  = & e_{q} \frac{N^2 g^2 N_c}{ P^+} 
\int \frac{d^{2} k_{\perp} d x}{
2(2 \pi)^3 x } \bigg[ -4 \Big( f_1 x P^+  
-x P^+ k^{2}_{\perp} - 2 f_1  P^+  + 
2 k^{2}_{\perp} P^{+}  - \frac{x P^+ q^2}{4} \Big) \nonumber \\ 
& & - \frac{4 f_1 P^{+}}{x} + 8 P^{+} (x - 1) m_{q} m_{\bar{q}} - 
4 x P^+ m_{q}^2 \bigg]
\theta(x) \theta(1-x) \ \Phi^*_f(x,k_{\perp}) 
\Phi_i(x,k_{\perp})  \ . \nonumber \\ 
\ F^{+}_{\bar{q}}(q^2) & = & q \ \leftrightarrow \ \bar{q}  \  \mbox{in}  
\ F^{+}_{q}(q^2)  \  \ ,   
\label{form} 
\end{eqnarray}
where $f_1=k_{\perp}^{2}+m^2_{\bar q}$,   
$f_2=(P - k)_{\perp}^{2}+m^2_{ q}$,   
$f_3=(P^{\prime } - k)_{\perp}^{2}+m^2_{ q}$,   
$f_4=(P-k)_{\perp}^{2}+M^2_{R}$ and 
$f_5=(P^{\prime}-k)_{\perp}^{2}+M^2_{R}$. 
The light-front wave function for 
the kaon with the nonsymmetric vertex is written like: 
\begin{equation}
\Phi_q^i(x,k_{\perp})=
\biggl[ 
\frac{N}{(1-x)^2  
(m_{K^+}^2 - M_0^2) (m_{K^+}^2 - M_R^2)}
\biggr] \ ,
\label{wavefunction}
\end{equation}
where $x=k^{+}/P^{+}$ is the momentum fraction carried 
by the quark, $0<x<1$ and $M_R^2$ is the square mass operator  
\begin{equation} 
M^2_{R}=M^2(m_{\bar{q}},M_R) = 
\frac{k^2_\perp+m^2_{\bar q}}{x}+ 
\frac{(P-k)^2_\perp+M^2_{R}}{1-x}-P^2_\perp \ .
\end{equation}  
The free quark square mass operator is given by 
$M^2_0 = M^2(m_{q}^2,m_{\bar q}^2)$. 
The normalization constant  $N$ is determined with the 
charge conservation condition $F^+_{K^+}(0)=1$. 
The calculation of the kaon e.m.~form factor in the light-front 
with $J^{+}_{K}$ gives the same result as a covariant calculation~\cite{Pereira}.
%%  \section{Results}
The parameters utilized in the case 
of the nonsymmetric vertex 
are the constituent quark masses $m_q=m_{u}=0.220$~GeV, 
$m_{\bar{q}}=m_{\bar{s}}=0.508$~GeV 
and the regulator mass $m_R=1.0$~GeV, which are adjusted 
to fit the charge radius of the kaon. 
The kaon mass considered in the calculations is the experimental value, 
$m_{K}=0.496$~GeV~\cite{PDG}.
\vspace{0.5cm}
\begin{figure} [h]
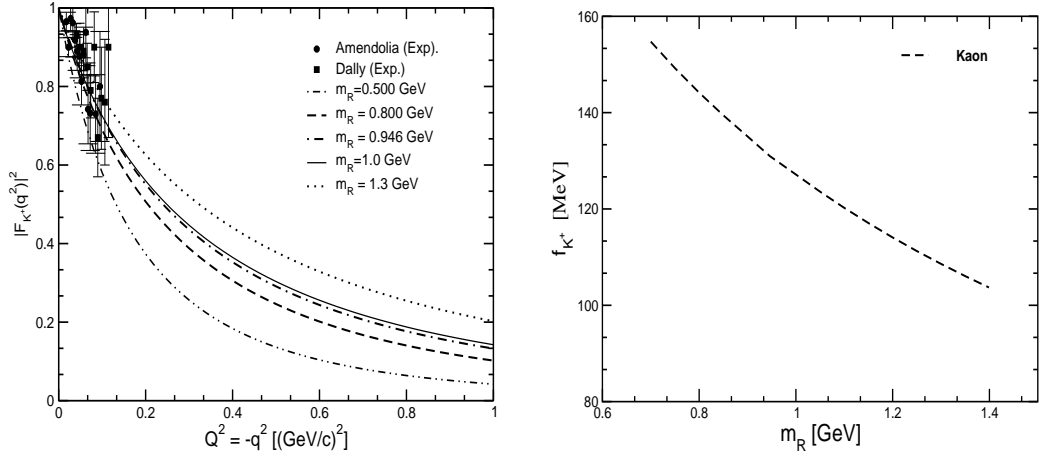
%% [tbh!]
\vspace{0.3cm}
\centerline{\epsfig{figure=pafig7.eps,width=6.5cm,height=6.0cm}
\hspace{0.5cm}
\epsfig{figure=kafig2.eps,width=6.5cm,height=6.5cm}} 
\caption{The
left frame shows the kaon e.m. form factor calculated with the 
model presented here for
different values of $m_R$ and compared with the experimental 
data~\cite{kaondata}. The results are covariant and free of the zero modes contributions~
(see the ref.~\cite{Pacheco2006} and references therein).  
The right frame shows the kaon decay constant 
 for different values of $m_R$.} %% compared.}
%% with the experimental data~\cite{PDG} 
\label{fig1}
\end{figure}

 \vspace*{0.2cm}

With these parameters, the calculated charge 
radius of the kaon and the leptonic decay constant are 
$< r_{K^+} > = 0.636~fm$ and $f_{K^+}=126.9$~MeV, 
close to the experimental values, 
$<r_{K^+}>^{exp} = 0.560~fm$ and  $f^{exp}_{K^+}=110.4$~MeV~\cite{kaondata}, and the e.m. 
form factor for the kaon is shown in the left frame of figure 1. Further, in the right frame of 
figure 1, we show the calculated decay constant as function of the regulator mass $m_R$.
In our calculations in such a light-front model, we obtain the same result of a covariant calculation, 
so that we really verify that $J_{K}^+$ does not have pair-term contributions for the kaon e.m.~form factor. 
%% In the case of $J_{K}^-$, the light-front calculation 
%% gives results quite different from the covariant results. After the inclusion of 
%% the pair terms, with $J_{K}^-$, we observe a 
%% complete agreement between the light-front and covariant results. 
%% In conclusion, the $J_{K}^+$ component of the kaon e.c.~are obtained in the 
%% light-front with a costituent quark model. 
%% In the case of $J_{K}^-$, we note that the  
%% pair terms or non-valence terms are essential to obtain 
%% the full covariance and a complete agreement between the  covariant and 
%% the light-front results for the kaon e.m.~form factor. 
At very low momentum transfer, the light-front model presented 
here gives better agreement with experimental data
~\cite{PDG,kaondata}.
 
%%%%%% Falta um parágrafo para explicar as figuras apresentadas, senão não faz sentido colocá-las. 
 
%%  \section{Conclusion}
%% The light-front approach it apropriate to realize hadronic 
%% calcutions for bound state; however problems related with the broken of the 
%% rotational symmetry in the light-front approach are 
%% important and the pair terms contribution for the covariance restoration is also necessary. 
%% After the pair terms inclusion in the 
%% matrix elements of the electromagnetic current, the 
%% covariance is complete restorate and no matter with 
%% the component of the e.c.~are utilized 
%% in order to extract the kaon e.m.~form factor with the 
%% light-front approach. For the light-front model for the vertex utilized in the 
%% prresent work, the kaon observables are very well described. 

%% \begin{theacknowledgments}

\vspace{0.5cm}
{\bf Acknowledgments:}~We thank the brazilian agencies  
{\it CAPES,~Coordenadoria de Aperfei\c{c}omento 
de Pessoal de N\'\i vel Superior}, {\it CNPq,~Conselho Nacional de Desenvolvimento
Cient\'\i fico e Tecnol\'ogico}, 
 and {\it FAPESP,~Funda\c{c}\~ao de Amparo
\`a Pesquisa do Estado de S\~ao Paulo}, for financial support.
%%This work was supported by Brazilian agencies {\it FAPESP}, {\it CAPES} and {\it CNPq}.

%% \end{theacknowledgments}

\bibliographystyle{aipproc}   % if natbib is available

\end{document}